\documentstyle[12pt]{article}
\def \lim{\rm Lim}

\begin{document}
\baselineskip=24pt
\bibliographystyle{plain}


\setlength{\textwidth}{15.5cm}

\title{Scalar $\sigma$ meson via chiral and crossing dynamics}

\author{M.D.~Scadron\footnote{Permanent Address:~~Physics Department, 
University of Arizona, Tucson, AZ 85721, USA} \\ 
TRIUMF, 4004 Wesbrook Mall, Vancouver, BC, Canada, V6T 2A3}

\date{}

\maketitle

\begin{abstract}
We show that the non-strange scalar $\sigma$ meson, as now reported in the 1996
PDG tables, is a natural consequence of crossing symmetry as well as chiral
dynamics for both strong interaction low energy $\pi\pi$ scattering and also $K
\to 2\pi$ weak decays.
\end{abstract}


\section{Introduction}
The 1996 Particle Data Group (PDG) tables[1] now includes a broad non-strange
I=0 scalar $\sigma$ resonance referred to as $f_0$ (400-1200). This is based
in part on the T\"{o}rnqvist-Roos[2] re-analysis of low energy $\pi\pi$
scattering, finding a broad non-strange $\sigma$ meson in the 400-900~MeV
region with pole position $\sqrt{s_0}$ = 0.470 - i~0.250~MeV. Several
later comments in PRL[3-5] all stress the importance of rejecting[3] or
confirming[4,5] the above T\"{o}rnqvist-Roos[2] $\sigma$ meson analysis based on
(t-channel) crossing symmetry of this $\pi\pi$ process.

In this brief report we offer such a $\sigma$ meson-inspired crossing symmetry
model in support of Refs.~[2,4,5] based on chiral dynamics for strong
interaction $\pi\pi$ scattering (Sect.~II). This in turn supports the recent
$s$-wave $\pi\pi$ phase shift analyses[6] in Sect.~III using a negative
background phase obtaining a broad $\sigma$ resonance in the 535-650~MeV mass
region. This is more in line with the prior analysis of Ref.~[5] and with the
dynamically generated quark-level linear $\sigma$ model (L$\sigma$M) theory of
Ref.~[7] predicting $m_\sigma \approx$ 650~MeV. Section~IV looks instead at
processes involving two final-state pions where crossing symmetry plays no
role, such as for the DM2 experiment[8] J/$\Psi \to \omega\pi\pi$ and 
for $\pi$N $\to \pi\pi$N polarization measurements[9]. Section~V
extends the prior crossing-symmetric strong interaction chiral dynamics to the
non-leptonic weak interaction $\Delta$I = $\frac{1}{2}$ decays $K^\circ
\to 2\pi$. We give our conclusions in Sect.~VI.

\section{Strong Interactions, Crossing Symmetry and the $\sigma$ Meson}
It has long been understood[10-12] that the non-strange isospin I=0 $\sigma$ 
meson is
the chiral partner of the I=1 pion. In fact Gell-Mann-L\'{e}vy's[10,11]
nucleon-level L$\sigma$M requires the meson-meson couplings to satisfy (with
$f_\pi \approx$ 93~MeV)
\begin{eqnarray}
g_{\sigma\pi\pi} = \frac{m_\sigma^2 - m_\pi^2}{2 f_\pi} = \lambda f_\pi\,\, ,
\end{eqnarray}
where $g_{\sigma\pi\pi}$ and $\lambda$ are the cubic and quartic meson couplings
respectively. On the other hand, the $\sigma$ meson pole for the $\pi\pi$
scattering amplitude at the soft point $s=m_\pi^2$ using (1) becomes
\begin{eqnarray}
M_{\pi\pi}^{\sigma {\rm pole}} = \frac{2g_{\sigma\pi\pi}^2}{s-m_\sigma^2} 
\to 
\frac{2 g_{\sigma\pi\pi}^2}{m_\pi^2 - 
m_\sigma^2} = -\lambda = -M_{\pi\pi}^{\rm contact}\,\, .
\end{eqnarray}

The complete tree-level L$\sigma$M $\pi\pi$ amplitude is the sum of the quartic
contact amplitude $\lambda$ plus $\sigma$ poles added in a {\em crossing
symmetric} fashion from the s, t and u-channels. Using the chiral symmetry
soft-pion limit (2) combined with the (non-soft) Mandelstam relation s~+~t~+~
u~=~4$m_\pi^2$, the lead $\lambda$ contact $\pi\pi$ amplitude miraculously
cancels[11]. Not surprisingly, the resulting net $\pi^a\pi^b \to
\pi^c\pi^d$ amplitude in the L$\sigma$M is the low energy model-independent
Weinberg amplitude[13].
\begin{eqnarray}
M_{\pi\pi} = \frac{s - m_\pi^2}{f_\pi^2}~\delta^{ab} \delta^{cd} + \frac{t -
m_\pi^2}{f_\pi^2}~\delta^{ac} \delta^{bd} + \frac{u - m_\pi^2}{f_\pi^2}~
\delta^{ad} \delta^{bc}\,\, ,
\end{eqnarray}
due to partial conservation of axial currents (PCAC) applied
crossing-consistently to all three s, t, u-channels. Recall that the underlying
PCAC identity $\partial A^i = f_\pi m_\pi^2 \phi_\pi$, upon which the Weinberg
crossing-symmetric PCAC relation (3) is based, was originally obtained from the
L$\sigma$M lagrangian[10,11].

Although the above (L$\sigma$M) Weinberg PCAC $\pi\pi$ amplitude (3) predicts
an $s$-wave I=0 scattering length[13] $a_{\pi\pi}^{(0)} = 7 m_\pi/32\pi f_\pi^2
\approx 0.16~m_\pi^{-1}$ which is $\sim$30\% less than first obtained from
$K_{\ell 4}$ data[14], more precise experiments are now under consideration.
Moreover a simple chiral-breaking scattering-length correction $\Delta
a_{\pi\pi}^0$ follows from the L$\sigma$M using a Weinberg-like
crossing-symmetric form[15]
\begin{eqnarray}
M_{\pi\pi}^{abcd} = A(s,t,u) \delta^{ab} \delta^{cd} + A(t,s,u) \delta^{ac}
\delta^{bd} + A(u,t,s) \delta^{ad} \delta^{bc}\,\, , \\
A^{L \sigma M} (s,t,u) = -2\lambda \left[1 - \frac{2 \lambda 
f_\pi^2}{m_\sigma^2 -s}\right]~=~\left(\frac{m_\sigma^2 - m_\pi^2}
{m_\sigma^2 - s}\right) \left(\frac{s - m_\pi^2}{f_\pi^2}\right)\,\, ,
\end{eqnarray}
where the L$\sigma$M Eq.~(1) has been used to obtain the second form of (5).
Then the I=0 s-channel amplitude 3A(s, t, u) + A(t, s, u) + A(u, t, s) predicts
the $s$-wave scattering length at s = 4$m_\pi^2$, t = u = 0 using the L$\sigma$M
amplitude (5) with $\varepsilon = m_\pi^2/m_\sigma^2 \approx 0.046$ for the
L$\sigma$M mass[7] $m_\sigma \approx$ 650~MeV:
\begin{eqnarray}
a_{\pi\pi}^{(0)}|_{\rm{L}\sigma {\rm M}} \approx \left( \frac{7 + \varepsilon}{1 -
4\varepsilon} \right) \frac{m_\pi}{32\pi f_\pi^2} \approx (1.23) \frac{7
m_\pi}{32\pi f_\pi^2} \approx 0.20 m_\pi^{-1}\,\, .
\end{eqnarray}
This simple 23\% L$\sigma$M enhancement of the Weinberg PCAC prediction[13]
agrees in magnitude with the much more complicated one-loop order chiral
perturbation theory approach[16] which also predicts an $s$-wave scattering
length correction of order $\Delta a_{\pi\pi}^0 \sim 0.04 m_\pi^{-1}$. This
indirectly supports a $\sigma$(650) scalar meson mass scale as used in (6).

The above ``miraculous (chiral symmetry) cancellation", due to Eqs.~(1) and
(2) has been extended to final-state pionic processes $A_1 \to
\pi(\pi\pi)_{s{\rm-wave}}$[17], $\gamma\gamma \to 2\pi^0$[18] and $\pi^- p
\to \pi^-\pi^+n$. In all of these cases the above L$\sigma$M
``miraculous cancellation" is simulated by a (non-strange) quark box -- quark
triangle cancellation due to the Dirac-matrix identity[17,18]
\begin{eqnarray}
\frac{1}{\gamma.p-m}~2m\gamma_5~\frac{1}{\gamma.p-m}~=~-\gamma_5~
\frac{1}{\gamma.p-m}~-~\frac{1}{\gamma.p-m}~\gamma_5\,\, ,
\end{eqnarray}
combined with the quark-level Goldberger relation (GTR) $f_\pi g_{\pi qq} = 
m_q$ and the L$\sigma$M meson couplings in (1).

Then the $u,d$ quark box graph in Fig.~1a for $A_1 \to 3\pi$ in the
chiral limit (miraculously) cancels the quark triangle graph of Fig.~1b coupled
to the $\sigma$ meson because of the GTR and the L$\sigma$M chiral meson 
identity (1) along with the minus signs on the right-hand-side (rhs) of (7):
\begin{eqnarray}
M_{A_13\pi}^{\rm box} + M_{A_13\pi}^{\rm tri} \to -\frac{1}{f_\pi} M(A_1
\to \sigma\pi) + \frac{1}{f_\pi} M(A_1 \to \sigma\pi) = 0\,\, .
\end{eqnarray}
This soft pion theorem[17] in (8) is compatible with the PDG tables[1] listing
the decay rate $\Gamma[A_1 \to \pi(\pi\pi)_{sw}] = 1 \pm 1$~MeV.

Similarly, the $\gamma\gamma \to 2\pi^0$ quark box graph suppresses the
quark triangle $\sigma$ resonance graph in the 700~MeV region, also compatible
with $\gamma\gamma \to 2\pi^0$ cross section data[18]. Finally, the
peripheral pion in $\pi^- p \to \pi^-\pi^+n$ sets up an analogous
$\pi\pi$ or quark box -- quark triangle $s$-wave soft pion cancellation which
completely suppresses any such $\sigma$ resonance -- also an experimental fact
for $\pi^- p \to \pi^-\pi^+n$.

\section{$\pi\pi$ Phase Shifts}
The above miraculous (chiral) cancellation in $\pi\pi \to \pi\pi, A_1
\to 3\pi, \gamma\gamma \to 2\pi^0$ and $\pi^- p \to
\pi^-\pi^+n$ amplitudes and in data lends indirect support to the analyses of
Refs.~[2,4,5]. Reference~[3] claims instead that the I=0 and I=2 $\pi\pi$ phase
shifts require t-channel forces due to ``exotic", crossing-asymmetric
resonances in the I=$\frac{3}{2}$ and 2 cross-channels rather than due a broad
low-mass scalar $\sigma$ meson (in the s-channel). We suggest that this latter
picture in Ref.~[3] does not take account of the crossing-symmetric extent of
the chiral $\pi\pi$ forces in all three s, t and u-channels, leading to the
above miraculous chiral cancellation.

Specifically the recent $\pi\pi$ phase shift analyses in Refs.~[6] use a
negative background phase approach compatible with unitarity. This background
phase has a hard core of size $r_c \approx 0.63~fm$ (the pion charged radius)
such that $\delta^{BG} = -p_\pi^{\rm CM}r_c$. Combining this background phase with
the observed $\pi\pi$ phase shifts (e.g., of CERN-Munich or Cason {\em et
al.}), the new I=0 phase shift goes through 90$^\circ$ resonance in the range
535-650~MeV, while the I=2 phase shift does not resonate but remains negative
as observed. References~[6] justify this background phase approach because of
the ``compensating $\lambda\phi^4$ contact (L$\sigma$M) interaction". From our
Sect.~II we rephrase this as due to the {\em crossing symmetric}
L$\sigma$M chiral `miraculous cancellation'[11] which recovers Weinberg's[13]
PCAC $\pi\pi$ amplitude in our Eq.~(3). 

Then Refs.~[6] choose a slightly model-dependent form factor F(s) (designed to
fit the lower energy region below 400~MeV) along with the best-fitted $\sigma
\to \pi\pi$ effective coupling (double the L$\sigma$M field theory
coupling (1)). This gives the resonant $\sigma$ width[6]
\begin{eqnarray}
\Gamma_R(s) = \frac{p_\pi^{\rm CM}}{8\pi s} [g_R F(s)]^2 \approx 340~
{\rm MeV} \ \ {\rm
at}~~~\sqrt{s}_R \approx 600~{\rm MeV},~~~g_R \approx 3.6~{\rm GeV} \,\, ,
\end{eqnarray}
for $p_\pi^{\rm CM} = \sqrt{s/4 - m_\pi^2} \approx 260$~MeV. However, the
decay width in (9) accounts only for $\sigma \to \pi^+\pi^-$ decay. To
include as well the $\sigma \to \pi^0\pi^0$ decay mode, one must scale 
up (9) by a factor of 3/2:
\begin{eqnarray}
\Gamma_{\sigma \to 2\pi} = \frac{3}{2} \Gamma_R(s) \approx 510~{\rm MeV} \,\, ,
\end{eqnarray}
not incompatible with Refs.~[1,2,5] but still slightly below Weinberg's recent
mended chiral symmetry (MCS) prediction[19]
\begin{eqnarray}
\global\def\theequation{11a}
\Gamma_{\sigma \to 2\pi}^{\rm MCS} = \frac{9}{2} \Gamma_\rho \approx
680~{\rm MeV} \,\, ,
\end{eqnarray}
or the L$\sigma$M decay width[15]
\begin{eqnarray}
\global\def\theequation{11b}
\Gamma_{\sigma \to 2\pi}^{L\sigma M} = \frac{3}{2}~~  
\frac{p_\pi^{\rm CM}}{8\pi}~~\frac{(2g_{\sigma\pi\pi})^2}{m_\sigma^2} \approx
580~{\rm MeV}\,\, ,
\end{eqnarray}
for $m_\sigma \approx$ 600~MeV.
Note too that the best fit $\sigma \to \pi^+\pi^-$ effective coupling
in Refs.~[6] of 3.60~GeV is close to the L$\sigma$M value in (1) at $m_\sigma^R
\approx 600$~MeV:
\begin{eqnarray}
\global\def\theequation{\arabic{equation}}
\setcounter{equation}{12}
g_R \to 2 g_{\sigma\pi\pi} = (m_\sigma^2 - m_\pi^2)/f_\pi \approx
3.66~{\rm GeV} \,\, .
\end{eqnarray}

\section{Crossing-Asymmetric Determinations of $\sigma$ (600-750)} 
With hindsight, the clearest way to measure the $\sigma \to \pi\pi$ 
signal is to avoid
$\pi\pi \to \pi\pi, \gamma\gamma \to 2\pi^\circ, \pi^- p
\to \pi^-\pi^+n$ scatterings or $A_1 \to \pi(\pi\pi)_{sw}$
decay, since these processes are always plagued by the $\pi\pi$ miraculous
chiral cancellation in (2) or an underlying quark box -- triangle cancellation
due to (7) as in (8). First consider the 1989 DM2 experiment[8] $J/\Psi
\to \omega\pi\pi$. Their Fig.~13 fits of the $\pi^+\pi^-$ and
$\pi^\circ\pi^\circ$ distributions clearly show the known non-strange narrow
$f_2$(1270) resonance along with a broad $\sigma$(500) ``bump" (both bumps are
non-strange and the accompanying $\omega$ is 97\% non-strange). Moreover, 
DM2 measured the
(low mass) $\sigma$ width as[8]
\begin{eqnarray}
\Gamma_{\sigma \to \pi\pi}^{\rm DM2} = 494 \pm 58~{\rm MeV} \,\, ,
\end{eqnarray}
very close to the modified Ref.~[6] $\sigma$ width fit of 510~MeV in Eq.~(10).

Finally, this Fig.~13 of DM2[8] clearly shows that the nearby $f_0$(980) bump
in the $\pi\pi$ distribution is only a ``pimple" by comparison. This suggests
that the observed[1] $f_0(980) \to \pi\pi$ decay mode proceeds via a
small $\sigma - f_0$ mixing angle and that $f_0$(980) is primarily an
$\overline{s}s$ meson, compatible with the analyses of Refs.~[2,20]. However,
such a conclusion is not compatible with the $\overline{q}\overline{q}qq$ or
$K\overline{K}$ molecule studies noted in Ref.~[3].

Lastly, polarization measurements are also immune to the (spinless) miraculous
chiral cancellation[11] in $\pi\pi \to \pi\pi$. This detailed
polarization analysis of Ref.~[9] approximately obtains the $\rho$ (770) mass 
and 150~MeV decay width. While the resulting
$\sigma$ mass of 750~MeV is well within the range reported in the 1996 PDG[1]
and closer to the $\sigma$ mass earlier extracted from $\pi\pi \to
K\overline{K}$ studies in Ref.~[21], the inferred $\sigma$ width of
$\Gamma_\sigma \sim 200-300$~MeV in Ref.~[9] is much narrower than reported in
Refs.~[1, 2, 8, 21] or in our above analysis.

\section{$K^\circ \to 2\pi$ Weak Decays and the $\sigma$(600-700)
Meson}
To show that the $\sigma$(600-700) scalar meson also arises with chiral
crossing-symmetric weak forces, we consider the $\Delta$I=1/{2} --
dominant $K^\circ \to 2\pi$ decays. To manifest such a 
$\Delta$I=1/2 transition, we first consider the virtual $K^\circ~I=\frac{1}{2}$
meson t-channel tadpole graph of Fig.~2. Here the weak tadpole transition
$<0|H_w|K^\circ>$ clearly selects out the $\Delta$I=1/2 part of the
parity-violating component of $H_w$, while the adjoining strong interaction
$K^\circ\overline{K}^\circ \to \pi\pi$ is the kaon analogue of the 
t-channel
$\pi\pi \to \pi\pi$, with Weinberg-type PCAC[22] amplitude
($t-m_\pi^2)/2f_\pi^2$ for $t=(p_K-0)^2=m_K^2$. Then the $\Delta$I=1/2
amplitude magnitude is[23]
\begin{eqnarray}
|<\pi\pi|H_w|K^\circ>| = \frac{|<0|H_w|K^\circ>|}{2f_\pi^2}
(1-m_\pi^2/m_K^2)\,\, .
\end{eqnarray}

A crossed version of this $\Delta$I=1/2 transition (14) is due to the
s-channel I=0 $\sigma$ meson pole graph of Fig.~3 at $s=m_K^2$[24]. This leads
to the $\Delta$I=1/2 amplitude magnitude
\begin{eqnarray}
\global\def\theequation{15a}
|<\pi\pi|H_w|K^\circ>|=|<\pi\pi|\sigma> \frac{1}{m_K^2-m_\sigma^2 +
im_\sigma\Gamma_\sigma} <\sigma|H_w|K^\circ>|\,\, .
\end{eqnarray}
Applying chiral symmetry $<\sigma|H_w|K^\circ>=<\pi^\circ|H_w|K^\circ>$
(converting the former parity-violating to the latter parity-conserving
transition) along with the L$\sigma$M values $|<\pi\pi|\sigma>|\linebreak[4] 
= m_\sigma^2/f_\pi$
from (1) and $\Gamma_\sigma \approx m_\sigma$ to (15a), one sees that the
$\sigma$ mass scale cancels out of (15a), yielding[25]
\begin{eqnarray}
\global\def\theequation{15b}
|<\pi\pi|H_w|K^\circ>| \approx |<\pi^\circ|H_w|K^\circ>/f_\pi|\,\, .
\end{eqnarray}

Not only has (15b) been derived by other chiral methods[26], but (15b) also is
equivalent to (14) in the $m_\pi=0$ chiral limit because weak chirality
$[Q,H_w]=-[Q_5,H_w]$ for V-A weak currents and PCAC clearly require
$|<\pi^\circ|H_w|K^\circ>| \approx |<0|H_w|K^\circ>/2f_\pi|$, as needed.

Thus, we see that the existence of an I=0 scalar $\sigma$ meson below 1~GeV
manifests crossing symmetry (from the t to the s-channel) for the dominant
$\Delta$I=1/2 equivalent amplitudes (14) and (15b). Further use of the
quark model and the GIM mechanism[27] converts the $K_{2\pi}^\circ$ amplitudes
in (14) or (15b) to the scale[23] $24 \times 10^{-8}$~GeV, close to the
observed $K_{2\pi}^\circ$ amplitudes[1].

While the $\Delta$I=1/2 $K^\circ \to 2\pi$ decays are controlled by the
tadpole diagram in Fig.~2 (similar to $\Delta$I=1 Coleman-Glashow tadpole for
electromagnetic (em) mass splittings[28,29]), the smaller  $\Delta$I=3/2 $K^+
\to 2\pi$ amplitude is in fact suppressed by ``exotic"  I=3/2 meson
cross-channel Regge trajectories[30] (in a manner similar to the I=2
cross-channel exotic Regge exchange for the $\pi^+ - \pi^\circ$ em mass
difference[31]). This latter duality nature of crossing symmetry for  exotic
I=3/2 and I=2 channels was invoked in Ref.~[3] to {\em reject} the low mass
$\sigma$ meson scheme reported in the 1996 PDG tables~[1] based in part on the
data analysis of Ref.~[2]. That is, for exotic I=2 and I=3/2 (t-channel) dual
exchanges, the dynamical dispersion relations thus generated are unsubtracted,
so that one can then directly estimate the observed $\Delta$I=2 em mass
differences[32] and also the $\Delta$I=3/2 weak $K_{2\pi}^+$ decay
amplitude[33]. However, for I=1 and I=1/2 dual exchanges, the resulting
dispersion relations are once-subtracted, with subtraction constants
corresponding to contact  $\Delta$I=1 and $\Delta$I=1/2 tadpole diagrams for em
and weak transitions, respectively. Contrary to Ref.~[3], we instead suggest
that these duality pictures for exotic I=3/2 and I=2 channels of Refs.~[30,31]
in fact help {\em support} the existence of the I=0 chiral $\sigma$ meson in
Refs.~[2,4-7].

\section{Summary}
We have studied both strong and weak interactions involving two final-state
pions at low energy, using chiral and crossing symmetry to reaffirm the
existence of the low-mass I=0 scalar $\sigma$ meson below 1~GeV. This supports
the recent phenomenological data analyses in Refs.~[2,4-6] and the quark-level
linear $\sigma$ model [L$\sigma$M] theory of Ref.~[7].

In Sect.~II we focussed on $\pi\pi$ scattering and the crossing symmetry
miraculous chiral cancellation[11] in the L$\sigma$M and its extension to the
quark box -- quark triangle soft pion cancellation[17,18]. Such chiral
cancellations in $\pi\pi \to \pi\pi, A_1 \to 3\pi, \gamma\gamma
\to 2\pi^0, \pi^- p \to \pi^-\pi^+n$ in turn suppress the
appearance of the $\sigma$(600-700) meson. Then in Sect.~III we supported the
recent re-analyses[6] of $\pi\pi$ phase shift data invoking a negative
background phase. This led to an I=0 $\sigma$ meson in the 535-650~MeV region,
but with a broader width $\Gamma_\sigma \sim$ 500~MeV than found in Refs.~[6]
(but not incompatible with the 1996 PDG $\sigma$ width[1]).

In Sect.~IV we briefly reviewed two different crossing-asymmetric
determinations of the I=0~$\sigma$(600-750) which circumvent the above
crossing-symmetric `miraculous' chiral suppression of the $\sigma$ meson.
Finally, in Sect.~V we reviewed how the low mass I=0 $\sigma$ meson s-channel
pole for $\Delta$I=1/2 $K^0 \to 2\pi$ decays is needed to cross over to
the t-channel $\Delta$I=1/2 tadpole graph (which in turn fits data). This
$\Delta$I=1/2 crossing-symmetry $K \to \pi\pi$ picture was also
extended by crossing duality to justify why the (much smaller) $\Delta$I=3/2
$K_{2\pi}^+$ decay is controlled by exotic I=3/2 t-channel Regge
trajectories[30], while the above I=1/2 dispersion relation has a (tadpole)
non-exotic Regge subtraction constant.

\section{Acknowledgements}
The author is grateful for hospitality and partial support at TRIUMF.

\newpage

\section{References}

\begin{enumerate}
\item Particle Data Group, R.M.~Barnett {\em et al}., Phys. Rev. D
\underline{54}, Part I, 1 (1996).
\item N.A.~T\"{o}rnqvist and M.~Roos, Phys. Rev. Lett. \underline{76}, 1575 
(1996).
\item N.~Isgur and J.~Speth, Phys. Rev. Lett. \underline{77}, 2332 (1996) and
references therein.
\item N.A.~T\"{o}rnqvist and M.~Roos, Phys. Rev. Lett. \underline{77}, 2333 
(1996); \underline{78}, 1604 (1997).
\item M.~Harada, F.~Sannio and J.~Schechter, Phys. Rev. Lett. \underline{78},
1603 (1997) and references therein.
\item S.~Ishida, M.Y.~Ishida, H.~Takahashi, T.~Ishida, K.~Takamatsu and
T.~Tsuru, Prog. Theor. Phys. \underline{95}, 745 (1996); S.~Ishida, T.~Ishida,
M.~Ishida, K.~Takamatsu and T.~Tsuru, Prog. Theor. Phys. in press and
hep-ph/9610359.
\item R.~Delbourgo and M.D.~Scadron, Mod. Phys. Lett. A \underline{10}, 251
(1995).
\item DM2 Collab. J.~Augustin {\em et al}., Nucl. Phys. B \underline{320}, 1
(1989).
\item M.~Svec, Phys. Rev. D \underline{53}, 2343 (1996).
\item M.~Gell-Mann and M.~L\'{e}vy, Nuovo Cimento \underline{16}, 705 (1960).
\item V.~De~Alfaro, S.~Fubini, G.~Furlan and C.~Rossetti, Currents in Hadron
Physics (North Holland, 1973) pp. 324-327.
\item Y.~Nambu and G.~Jona-Lasinio, Phys. Rev. \underline{122}, 345 (1961).
\item S.~Weinberg, Phys. Rev. Lett. \underline{17}, 616 (1966).
\item L.~Rosselet {\em et al}., Phys. Rev. D \underline{15}, 574 (1977);
A.A.~Belkov and S.A.~Bunyatov, Sov. J. Nucl. Phys. \underline{29}, 666 (1979);
\underline{33}, 410 (1981).
\item See e.g., P.~Ko and S.~Rudaz, Phys. Rev. D \underline{50}, 6877 (1994).
\item J.~Gasser and H.~Leutwyler, Phys. Lett. B \underline{125}, 321, 325
(1983).
\item A.N.~Ivanov, M.~Nagy and M.D.~Scadron, Phys. Lett. B \underline{273}, 137
(1991).
\item A.N.~Ivanov, M.~Nagy and N.I.~Troitskaya, Mod. Phys. Lett. A
\underline{7}, 1997 (1992); M.D.~Scadron, Phys. At. Nucl. \underline{56}, 1595
(1993).
\item S.~Weinberg, Phys. Rev. Lett. \underline{65}, 1177 (1990).
\item N.A.~T\"{o}rnqvist, Z. Phys. C \underline{68}, 647 (1995).
\item P.~Estabrooks, Phys. Rev. D \underline{19}, 2678 (1979).
\item H.~Osborn, Nucl. Phys. B \underline{15}, 50 (1970); L.F.~Liu and
H.~Pagels, Phys. Rev. D \underline{5}, 1507 (1972).
\item See e.g., S.R.~Choudhury and M.D.~Scadron, Phys. Rev. D \underline{53},
2421 (1996) app. B.
\item T.~Morozumi, C.S.~Lim and A.I.~Sanda, Phys. Rev. Lett. \underline{65},
404 (1990); U.G.~Meissner, Comm. Nucl. Part. Phys. \underline{20}, 119 (1991).
\item R.E.~Karlsen and M.D.~Scadron, Mod. Phys. Lett. A \underline{6}, 543
(1991).
\item See e.g., R.E.~Karlsen and M.D.~Scadron, Phys. Rev. D \underline{45},
4108 (1992).
\item S.~Glashow, J.~Iliopoulos and L.~Maiani, Phys. Rev. D \underline{2}, 1285
(1970).
\item S.~Coleman and S.~Glashow, Phys. Rev. \underline{134}, B671 (1964).
\item S.A.~Coon and M.D.~Scadron, Phys. Rev. C \underline{51}, 2923 (1995) and
references therein.
\item G.~Nardulli, G.~Preparata and D.~Rotondi, Phys. Rev. D \underline{27},
557 (1983).
\item H.~Harari, Phys. Rev. Lett. \underline{17}, 1303 (1966); M.~Elitzur and
H.~Harari, Ann. Phys. (N.Y.) \underline{56}, 81 (1970).
\item See e.g., T.~Das, G.S.~Guralnik, V.S.~Mathur, F.E.~Low and J.E.~Young,
Phys. Rev. Lett. \underline{18}, 759 (1967); I.S.~Gerstein, B.W.~Lee, H.T.~Nieh
and H.J.~Schnitzer, ibid. \underline{19}, 1064 (1967).
\item See e.g., M.D. Scadron, Phys. Rev. D \underline{29}, 1375 (1984);
R.E.~Karlsen and M.D.~Scadron, ibid. D \underline{44}, 2192 (1991).
\end{enumerate}

\newpage

\section{Figure Captions}

\begin{itemize}
\item[Fig.~1]~~Quark box (a) and quark triangle (b) graphs for $A_1 \to
3\pi$.

\item[Fig.~2]~~$\Delta$I=1/2 t-channel $K^\circ$ tadpole graph for $K^\circ
\to 2\pi$. 

\item[Fig.~3]~~$\Delta$I=1/2 s-channel $\sigma$ pole graph for $K^\circ
\to 2\pi$.
\end{itemize}

\end{document}